\documentclass[preprintnumbers,amsmath,amssymb, prd,notitlepage,nofootinbib,superscriptaddress, showkeys,twocolumn]{revtex4-2}

\usepackage{hyperref}

\usepackage[T1]{fontenc}
\usepackage{aas_macros}
\usepackage{graphicx}
\usepackage{dcolumn}
\usepackage{bm}
\usepackage{lipsum, babel}
\usepackage{xcolor}
\definecolor{xlinkcolor}{cmyk}{1,1,0,0}
\usepackage{lineno}
\usepackage{orcidlink}
\usepackage{multirow}
\usepackage{graphicx}
\usepackage{dcolumn}
\usepackage{bm}
\usepackage{hyperref}
\usepackage{soul}
\hypersetup{colorlinks=true,linkcolor=xlinkcolor,citecolor=xlinkcolor,urlcolor=xlinkcolor}

\usepackage{siunitx}    
\DeclareSIUnit\parsec{pc}
\DeclareSIUnit\arcmin{$^\prime$}

\newcommand{\hun}{\,\mathrm{km}\,\mathrm{s}^{-1}\mathrm{Mpc}^{-1}}
\begin{document}

\preprint{APS/123-QED}

\title{\textbf{The Atacama Cosmology Telescope:\\ A demonstration of CMB lensing measurement from daytime data} 
}

\author{Irene Abril-Cabezas\orcidlink{0000-0003-3230-4589}}
\email{ia404@cam.ac.uk}
\affiliation{DAMTP, Centre for Mathematical Sciences, University of Cambridge, Wilberforce Road, Cambridge CB3 OWA, UK}
\affiliation{Kavli Institute for Cosmology Cambridge, Madingley Road, Cambridge, CB3 0HA, UK}

\author{Frank J. Qu\orcidlink{0000-0001-7805-1068}}
\affiliation{Kavli Institute for Particle Astrophysics and Cosmology, 382 Via Pueblo Mall, Stanford, CA 94305-4060, USA}
\affiliation{SLAC National Accelerator Laboratory, 2575 Sand Hill Road, Menlo Park, CA 94025, USA}

\author{Joshua~Kim\orcidlink{0000-0002-0935-3270}}
\affiliation{Department of Physics and Astronomy, University of Pennsylvania, 209 South 33rd Street, Philadelphia, PA 19104, USA}

\author{Mathew S. Madhavacheril\orcidlink{0000-0001-6740-5350}}
\affiliation{Department of Physics and Astronomy, University of Pennsylvania, 209 South 33rd Street, Philadelphia, PA 19104, USA}

\author{Karen Perez-Sarmiento\orcidlink{0009-0002-7452-2314}}
\affiliation{Department of Physics and Astronomy, University of Pennsylvania, 209 South 33rd Street, Philadelphia, PA 19104, USA}

\author{Zachary~Atkins\orcidlink{0000-0002-2287-1603}} 
\affiliation{Department of Physics and Astronomy, University of Pennsylvania, 209 South 33rd Street, Philadelphia, PA 19104, USA}
\affiliation{Joseph Henry Laboratories of Physics, Jadwin Hall, Princeton University, Princeton, NJ 08544, USA}

\author{Erminia Calabrese}
\affiliation{School of Physics and Astronomy, Cardiff University, The Parade, Cardiff, Wales CF24 3AA, UK}

\author{Anthony~Challinor\orcidlink{0000-0003-3479-7823}}
\affiliation{Institute of Astronomy, Madingley Road, Cambridge CB3 0HA, UK}
\affiliation{Kavli Institute for Cosmology Cambridge, Madingley Road, Cambridge, CB3 0HA, UK}
\affiliation{DAMTP, Centre for Mathematical Sciences, University of Cambridge, Wilberforce Road, Cambridge CB3 OWA, UK}

\author{Mark~J.~Devlin\,\orcidlink{0000-0002-3169-9761}}
\affiliation{Department of Physics and Astronomy, University of Pennsylvania, 209 South 33rd Street, Philadelphia, PA 19104, USA}

\author{Adriaan~J.~Duivenvoorden\orcidlink{0000-0003-2856-2382}}
\affiliation{Max-Planck-Institut fur Astrophysik, Karl-Schwarzschild-Str. 1, 85748 Garching, Germany}

\author{Jo~Dunkley\orcidlink{0000-0002-7450-2586}} \affiliation{Joseph Henry Laboratories of Physics, Jadwin Hall, Princeton University, Princeton, NJ 08544, USA}
\affiliation{Department of Astrophysical Sciences, Peyton Hall, Princeton University, Princeton, NJ 08544, USA}

\author{Alexander van Engelen}
\affiliation{School of Earth and Space Exploration, Arizona State University, Tempe, AZ 85287, USA}

\author{Simone Ferraro \orcidlink{0000-0003-4992-7854}}
\affiliation{Lawrence Berkeley National Laboratory, One Cyclotron Road, Berkeley, CA 94720, USA}
\affiliation{Berkeley Center for Cosmological Physics, Department of Physics, University of California, Berkeley, CA 94720, USA}

\author{Emily Finson}
\affiliation{Physics and Astronomy Department, Stony Brook University, Stony Brook, NY 11794, USA}

\author{Carlos Herv\'ias-Caimapo\orcidlink{0000-0002-4765-3426}}
\affiliation{Instituto de Astrof\'isica and Centro de Astro-Ingenier\'ia, Facultad de F\'isica, Pontificia Universidad Cat\'olica de Chile, Av. Vicu\~na Mackenna 4860, 7820436 Macul, Santiago, Chile}

\author{Matt Hilton\orcidlink{0000-0002-8490-8117}}
\affiliation{Wits Centre for Astrophysics, School of Physics, University of the Witwatersrand, Private Bag 3, 2050, Johannesburg, South Africa} \affiliation{Astrophysics Research Centre, School of Mathematics, Statistics, and Computer Science, University of KwaZulu-Natal, Westville Campus, Durban 4041, South Africa}

\author{Arthur Kosowsky\orcidlink{0000-0002-3734-331X}}
\affiliation{Department of Physics and Astronomy, University of Pittsburgh, Pittsburgh, PA 15260, USA}

\author{Aleksandra~Kusiak\orcidlink{0000-0002-1048-797}}
\affiliation{Institute of Astronomy, Madingley Road, Cambridge CB3 0HA, UK}
\affiliation{Kavli Institute for Cosmology Cambridge, Madingley Road, Cambridge, CB3 0HA, UK}

\author{Thibaut~Louis\orcidlink{0000-0002-6849-4217}} \affiliation{Universit\'e Paris-Saclay, CNRS/IN2P3, IJCLab, 91405 Orsay, France}

\author{Niall MacCrann\orcidlink{0000-0002-8998-3909}}
\affiliation{DAMTP, Centre for Mathematical Sciences, University of Cambridge, Wilberforce Road, Cambridge CB3 OWA, UK}
\affiliation{Kavli Institute for Cosmology Cambridge, Madingley Road, Cambridge, CB3 0HA, UK}

\author{Kavilan~Moodley\orcidlink{ 000-0001-6606-7142}}
\affiliation{Astrophysics Research Centre, School of Mathematics, Statistics, and Computer Science, University of KwaZulu-Natal, Westville Campus, Durban 4041, South Africa}

\author{Toshiya Namikawa\orcidlink{0000-0003-3070-9240}}
\affiliation{Center for Data-Driven Discovery, Kavli IPMU (WPI), UTIAS, The University of Tokyo, Kashiwa, 277-8583, Japan}
\affiliation{DAMTP, Centre for Mathematical Sciences, University of Cambridge, Wilberforce Road, Cambridge CB3 OWA, UK}
\affiliation{Kavli Institute for Cosmology Cambridge, Madingley Road, Cambridge, CB3 0HA, UK}

\author{Sigurd~N\ae ss\orcidlink{0000-0002-4478-7111}} 
\affiliation{Institute of Theoretical Astrophysics, University of Oslo, Norway}

\author{Lyman~A.~Page}
\affiliation{Joseph Henry Laboratories of Physics, Jadwin Hall,
Princeton University, Princeton, NJ 08544, USA}

\author{Adrien~La~Posta\orcidlink{0000-0002-2613-2445}} 
\affiliation{Department of Physics, University of Oxford, Keble Road, Oxford, OX1 3RH, UK}

\author{Emmanuel Schaan\orcidlink{0000-0002-4619-8927}}
\affiliation{Kavli Institute for Particle Astrophysics and Cosmology, 382 Via Pueblo Mall, Stanford, CA 94305-4060, USA}
\affiliation{SLAC National Accelerator Laboratory, 2575 Sand Hill Road, Menlo Park, CA 94025, USA}

\author{Neelima Sehgal\orcidlink{0000-0002-9674-4527}}
\affiliation{Physics and Astronomy Department, Stony Brook University, Stony Brook, NY 11794, USA}

\author{Blake D. Sherwin}
\affiliation{DAMTP, Centre for Mathematical Sciences, University of Cambridge, Wilberforce Road, Cambridge CB3 OWA, UK}
\affiliation{Kavli Institute for Cosmology Cambridge, Madingley Road, Cambridge, CB3 0HA, UK}

\author{Carlos E. Sierra\orcidlink{0000-0002-9246-5571}}
\affiliation{Kavli Institute for Particle Astrophysics and Cosmology, 382 Via Pueblo Mall, Stanford, CA 94305-4060, USA}
\affiliation{SLAC National Accelerator Laboratory, 2575 Sand Hill Road, Menlo Park, CA 94025, USA}

\author{Crist\'obal Sif\'on\orcidlink{0000-0002-8149-1352}}
\affiliation{Instituto de F\'isica, Pontificia Universidad Cat\'olica de Valpara\'iso, Casilla 4059, Valpara\'iso, Chile}

\author{Suzanne T. Staggs\orcidlink{0000-0002-7020-7301}} 
\affiliation{Joseph Henry Laboratories of Physics, Jadwin Hall, Princeton University, Princeton, NJ 08544, USA}

\author{Emilie~Storer\orcidlink{0000-0003-1592-9659}} \affiliation{Physics Department, McGill University, Montreal, QC H3A 0G4, Canada}
\affiliation{Joseph Henry Laboratories of Physics, Jadwin Hall, Princeton University, Princeton, NJ 08544, USA}

\author{Edward~J.~Wollack\orcidlink{0000-0002-7567-4451}}
\affiliation{NASA Goddard Space Flight Center, 8800 Greenbelt Road, Greenbelt, MD 20771, USA}

\date{\today}
             
\begin{abstract}
We present a cosmic microwave background (CMB) lensing power spectrum analysis using daytime data (11am--11pm UTC) gathered by the Atacama Cosmology Telescope (ACT) over the period 2017--2022 (ACT Data Release 6). This dataset is challenging to analyze because the Sun heats and deforms the telescope mirror, complicating the characterization of the telescope. We perform more than one hundred null and consistency checks to ensure the robustness of our measurement and its compatibility with nighttime observations. We detect the CMB lensing power spectrum at {18$\sigma$} significance, with an amplitude {$A_\textrm{lens} = 1.071 \pm 0.060$} with respect to the prediction from the best-fit \textit{Planck}-ACT CMB power spectrum $\Lambda$CDM cosmology. In combination with the Dark Energy Spectroscopic Instrument (DESI) Baryon Acoustic Oscillation (BAO) data, this corresponds to a constraint on the amplitude of matter fluctuations {$\sigma_8 = 0.842 \pm 0.026$}.  The analysis presented here is especially relevant for ground-based millimeter-wave CMB experiments at the Atacama site, paving the way for future analyses making use of both nighttime and daytime data to place tight constraints on cosmological parameters.
\end{abstract}

\maketitle


\section{\label{sec:intro}Introduction}As photons from the cosmic microwave background (CMB) travel to us, they encounter large-scale structure that gravitationally deflects their trajectories. This phenomenon is known as weak gravitational lensing of the CMB, and it has emerged as a robust and powerful observable for precision cosmology \citep[for a review, see, e.g.,][]{Lewis_2006}. Reference~\citep{Qu_2025} presented {the first joint likelihood-level combination of CMB lensing power spectrum measurements from multiple telescopes}, namely the Atacama Cosmology Telescope (ACT) \citep{Frank_ACT_lensing_2024, Madhavacheril_2024, MacCrann_2024}, \textit{Planck} \citep{Carron_2022} and the South Pole Telescope (SPT) \citep{Ge_2025}. It provided the most precise CMB lensing power spectrum measurement to date (with a signal-to-noise ratio SNR of 61), which can be used to probe the growth of structure and place tight constraints on neutrino mass \citep{Lesgourges_2006, Allison_2015} or primordial non-Gaussianity (in cross-correlation with another biased tracer) \citep{Schmittfull_2018, McCarthy_2023}.\medskip

\begin{figure*}
\includegraphics[width=0.85\textwidth,trim=2mm 2mm 2mm 2mm, clip]
{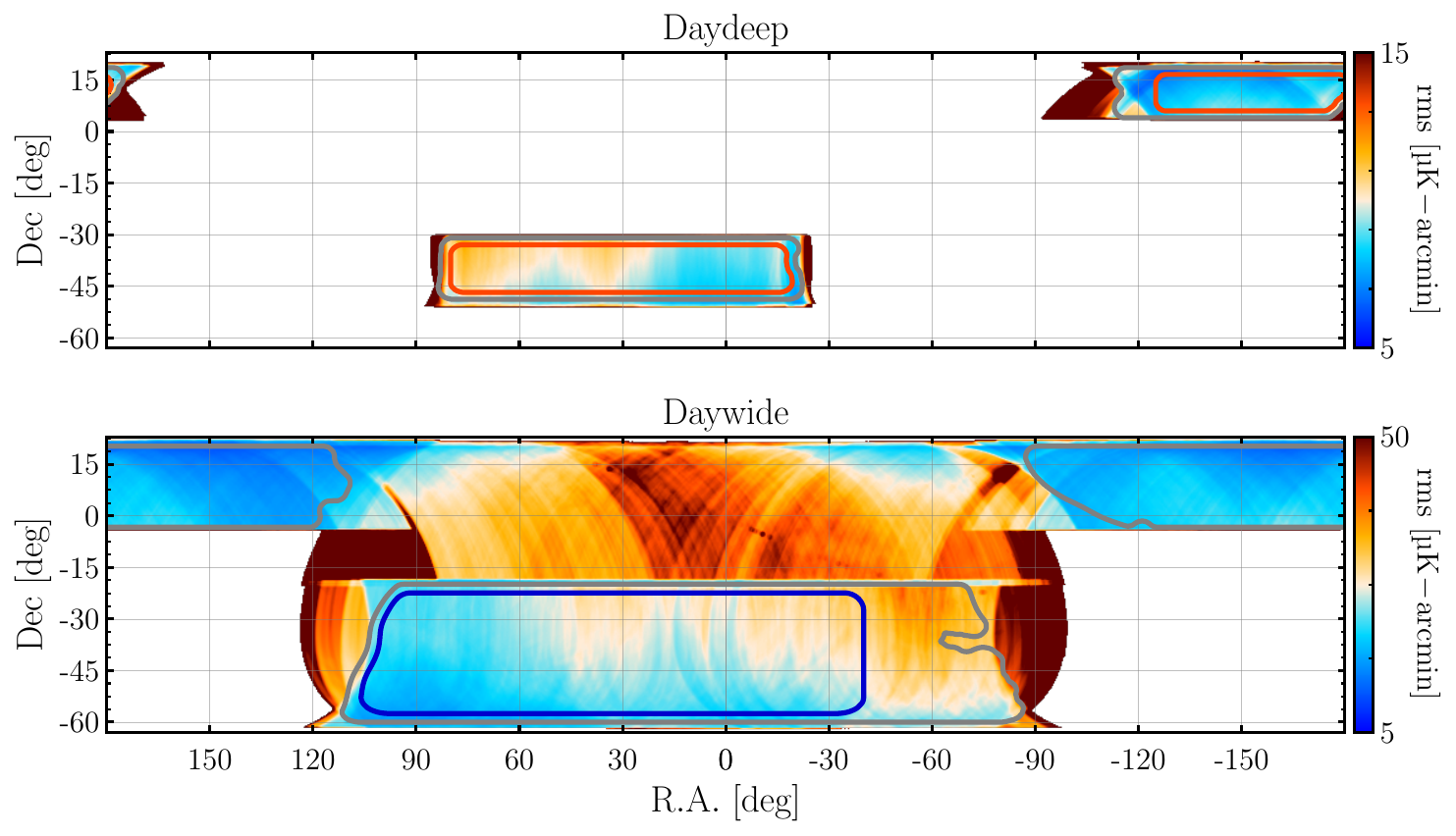}
\caption{\label{fig:rms_ivar} Map of root-mean-square (RMS) noise per pixel for \textit{daydeep} (\textit{top}) and \textit{daywide} (\textit{bottom})  derived from inverse-variance maps (that describe the noise behavior on small scales \citep{Atkins_2023, Naess_DR6_2025}). Regions such as the left half of the southern \textit{daydeep} patch are shallower due to data cuts prior to map-making, such as beam cuts (summarized in the beam-badness statistic; see \citep{Storer_thesis, Naess_DR6_2025}). The RMS maps, together with cross-linking and Galactic dust emission information, inform the choice of area that we model and simulate (encircled with gray contours). Further null tests inform the regions we use in our lensing power spectrum analysis: \textit{daywide} South, shown in blue, with a mean depth of $\SI{24}{\micro\kelvin}$-arcmin and effective sky fraction $f_\textrm{sky}=0.09$; and \textit{daydeep}, shown in orange, with a {mean depth of $\SI{9}{\micro\kelvin}$-arcmin and $f_\textrm{sky}=0.03$.} }
\end{figure*}

The ACT Data Release 6 (DR6) lensing analysis \citep{Frank_ACT_lensing_2024, Madhavacheril_2024, MacCrann_2024} utilized data taken during the night only. However, half of the time-ordered-data (TOD)\footnote{A TOD is a data unit, typically 11 minutes long \citep{Dunner_2013, Aiola_2020, Naess_DR6_2025}.} 
collected by ACT was during the day (between 11 a.m. and 11 p.m. UTC). This work focuses on the corresponding CMB lensing power spectrum analysis using daytime data only. Working with daytime data from ACT is challenging because the Sun heats and deforms the telescope structure and mirrors, making it difficult to characterize the beam profile \citep{Storer_thesis}.\footnote{It also leads to pointing offsets, although the new per-TOD pointing corrections used in the DR6 pipeline successfully correct for them \citep{Naess_DR6_2025}.}
We find two intuitive reasons why a measurement of CMB lensing could be robust against ACT daytime complexities. First, the time variation of the telescope beam on timescales of hours, which could lead to a smoothly varying position-dependent beam, would mostly affect the reconstruction of the largest-scale CMB lenses (which are already excluded from our analysis, see \citep{Frank_ACT_lensing_2024}). Second, beam mis-modeling would not produce the exact same effect on the CMB power spectrum as the squeezing and stretching due to real lensing, meaning that these two effects are not fully degenerate. With these arguments in mind, it is a promising avenue to utilize daytime data to increase the signal-to-noise of the current state-of-the-art ACT nighttime-only lensing measurement \citep[43$\sigma$ significance, ][]{Frank_ACT_lensing_2024}.\footnote{We note that the latest ACT-only lensing analysis \citep{Frank_ACT_lensing_2024,Madhavacheril_2024, MacCrann_2024} was based on nighttime observations using the \texttt{dr6.01} map version. Since then, we have made refinements to the map-making that improve the large-scale transfer function and polarization noise levels. This work uses daytime observations from the \texttt{dr6.02} re-processing of the ACT DR6 dataset that includes these improvements (ACT DR6 public data release).}\medskip

{We note that SPT and BICEP/Keck include data taken when the Sun is above the horizon but not near their field \citep[e.g.,][]{bicep_2016lensing, Ge_2025}. Because the Sun remains up for about six months at the South Pole, the resulting observational challenges (e.g., thermal stability) are quite different from those in the Atacama. In the Atacama, the \textsc{Polarbear} experiment included some daytime data in their lensing analysis \citep[e.g.,][]{polarbear_2020}, with a 24\% precision on the lensing amplitude for their full dataset.}\medskip

The inclusion of daytime data (this work) is one of the improvements anticipated for the final ACT analysis, which we denote ACT DR6+ \citep{Kim_2026, AbrilCabezas_plus_2026, Quplus_2026}.\medskip

\section{Methodology}

\subsection{\label{sec:dataset}Data collection}The final ACT DR6 dataset \citep{Naess_DR6_2025} was collected during the period 2017--2022. Until mid-2019, the daytime observations concentrated on two small regions,\footnote{These maps are labeled  \texttt{act\_dr6.02\_std\_DN\_day} and \texttt{act\_dr6.02\_std\_DS\_day} in the public ACT DR6 release. Details on data access are available in the \textit{Data Availability} section at the end of this work.}
which we refer to as \textit{daydeep}. They cover approximately 8\% of the sky with a median depth of $\SI{9}{\micro\kelvin}$-arcmin (see top panel of Fig.~\ref{fig:rms_ivar}). Thereafter, the daytime scanning strategy followed that of night. We refer to the corresponding observations as \textit{daywide}.\footnote{These maps are labeled  \texttt{act\_dr6.02\_std\_AA\_day} in the public ACT DR6 release.} 
They cover almost half the sky with a median depth of $\SI{27}{\micro\kelvin}$-arcmin (see bottom panel of Fig.~\ref{fig:rms_ivar}).\medskip

In this analysis, we only use data at \SI{90}{\giga\hertz} and \SI{150}{\giga\hertz}, which were gathered by ACT polarized arrays (PA) 5 and 6 \citep{Thornton_2016, Naess_DR6_2025} (thus excluding PA4 data, which includes measurements at \SI{220}{\giga\hertz}). This choice is motivated by the systematics and null tests carried out in previous ACT power spectrum and lensing analyses \citep{Frank_ACT_lensing_2024, Louis_DR6_2025}, which validated the corresponding nighttime data gathered by those polarized arrays, together with foreground contamination studies \citep{MacCrann_2024, AbrilCabezas_2025}. In addition, we only analyze PA5 data for \textit{daywide}; the PA6 dataset is too shallow because it was replaced with a low-frequency array (\SI{30}{\giga\hertz} and \SI{40}{\giga\hertz}) soon after the \textit{daywide} survey started. \medskip

\subsection{\label{sec:masks}Data selection and noise modeling}Our starting point is the ACT DR6 source-free maps.\footnote{These maps are labeled as \texttt{map\_srcfree} in the public ACT DR6 release.} 
Prior to the final making of those maps, various data cuts and corrections are applied (e.g., removing data samples with the Sun or Moon in the far sidelobes, or with high beam variability), described in more detail in  \cite{Storer_thesis, Naess_DR6_2025}. We further remove areas with poor cross-linking \citep{Atkins_2023} or with root-mean-square (RMS) map noise larger than $\SI{70}{\micro\kelvin}$-arcmin. Areas at the edge of the map with high noise are also removed after careful visual inspection. We further exclude regions of bright Galactic emission following the same approach as \citep{Frank_ACT_lensing_2024}, primarily by intersecting the ACT footprint with the \textit{Planck} 70\% Galactic mask.\footnote{\texttt{HFI\_Mask\_GalPlane-apo0\_2048\_R2.00.fits} is available on the Planck Legacy Archive (\url{https://pla.esac.esa.int/}). The 70\% mask retains the 70\% of the sky with the lowest dust contamination and has been validated to have a negligible impact on the inferred CMB lensing power spectrum amplitude for an ACT-like survey \citep{AbrilCabezas_2025}.}  
We use the common resulting area within \textit{daydeep}  or \textit{daywide} (shown with gray lines in Fig.~\ref{fig:rms_ivar}) to generate map-based noise models using the \textsc{mnms} approach~\citep{Atkins_2023}, and draw noise simulations from these. \medskip

For \textit{daydeep}, we further shrink the \textit{daydeep}  mask by $1.5^\circ$ around the perimeter to mitigate any edge effects. The corresponding effective sky fraction of \textit{daydeep}  after apodizing the masks on $3^\circ$ scales is $f_\textrm{sky}=3\%$. This is shown in orange in the top panel of Fig.~\ref{fig:rms_ivar}.
For \textit{daywide}, we shrink the \textit{daywide} mask by $2.5^\circ$ and exclude high variance regions when we perform the lensing reconstruction. The final mask is apodized on $3^\circ$ scales. In the end, we are forced to exclude completely the \textit{daywide} North region from our analysis as it fails multiple null tests.\footnote{These failures were not surprising to us since we found a two-point power-spectrum mismatch between the corresponding night and \textit{daywide} array-bands, especially at multipoles $\ell<1000$; however, an understanding of the physical mechanism was not identified and we thus decide not to include this region.} 
The final \textit{daywide} mask is marked in blue in the bottom panel of Fig.~\ref{fig:rms_ivar} ($f_\textrm{sky}=9\%$). \medskip

\subsection{\label{sec:calibration}Data pre-processing and calibration}Before the source-free maps pass through the lensing reconstruction pipeline,\footnote{\url{https://github.com/simonsobs/so-lenspipe}} 
we process them following the approach described in detail in \citep{Frank_ACT_lensing_2024}. In summary, this involves first downgrading the maps to a pixel resolution of $1\,\text{arcmin}$. Then, we inpaint regions around residual compact-object sources in our maps \citep{Frank_ACT_lensing_2024, Madhavacheril_2020} by smoothly interpolating from surrounding values.
The emission from tSZ-selected clusters \citep{DR6_clusters} is estimated and subtracted, for which we use the \textsc{nemo}\footnote{\url{https://nemo-sz.readthedocs.io/}}
software \citep{Hilton_2021}. {In particular, we select cluster candidates with signal-to-noise ratio $\bar{q} > 5.5$. We can only characterize reliably the tSZ emission from clusters located outside ``dusty regions'' (those where the \textit{Planck} \SI{353}{\giga\hertz} is above \SI{2}{\milli\kelvin} \citep{DR6_clusters}) and therefore inpaint all remaining ones.} We also deconvolve the pixel window function introduced by the downgrading operation and mask Fourier modes with $|\ell_x|<90$ and $|\ell_y|<50$ to remove ground pickup, following \citep{Louis_DR6_2025}. Finally, we beam-deconvolve each daytime map \citep{Duivenvoorden_2025}. \medskip

\begin{table}
\caption{\label{tab:cals}%
We find that daytime data can be made reliable for lensing with a suitable process of rescaling to nighttime data. We therefore apply an additional calibration factor (fourth column), which we treat as a gain, in the pre-processing of these maps. This factor is obtained by comparing each array-band daytime temperature power spectrum to the corresponding array-band power spectrum of nighttime observations, within the multipole range $1000<\ell<2000$. 
}
\begin{ruledtabular}
\begin{tabular}{cccc}
\textrm{Daytime}&
\textrm{Array}&
\textrm{Band [GHz]}&
\textrm{Calibration}\\
\colrule
\textit{daydeep}  & PA5 & 90 & 1.0376 \\
\textit{daydeep}  & PA5& 150 & 0.9705 \\ 
\textit{daydeep} &  PA6 & 90 & 1.0270\\
\textit{daydeep}  & PA6 & 150 & 0.9327\\ 
\textit{daywide} & PA5  & 90 & 1.0416\\
\textit{daywide} & PA5& 150 & 0.9788\\
\end{tabular}
\end{ruledtabular}
\end{table}

The daytime beams are estimated in \citep{Duivenvoorden_2025, Naess_DR6_2025} using both the profile of point sources in the daytime maps (which informs the small-scale behavior) and the relative CMB power spectrum between night and day maps for each array-band.\footnote{When this ratio is too noisy, both night and day spectra are calibrated with respect to \textit{Planck} first via their auto- and cross-spectra. Using \textit{Planck} as a noise-reducing template only introduces sensitivity to \textit{Planck}'s phases (but not to its amplitude).}
The beam derived from the CMB power spectrum comparison is valid for $\ell \lesssim 4000$, but has relatively large error bars throughout. The point source beam has much higher SNR but is only valid for $\ell > 2000$ due to low-pass filtering in its production. The point source beam is estimated from stacking all point sources within $10\times20~\SI{}{\deg\squared}$ tiles, which are then averaged to a common beam across all tiles using inverse-variance weighting. It has an unknown normalization because almost all the bright point sources are strongly variable. We merge these two beam measurements by fitting a relative normalization for the point source beam in their overlapping multipole range and fitting a smooth function\footnote{The model consists of a smooth function multiplying the nighttime beam. The smooth function is the product of two sigmoid functions, chosen \textit{ad-hoc} for its reasonably good fit to the data given its 6 degrees of freedom. The smoothing is necessary to reduce the scatter in what would otherwise be a quite noisy beam model, but the result still falls short of the reliability of the nighttime beam.}
to the normalized beams. The procedure results in an effective beam that also takes into account both the transfer function at low $\ell$ in total intensity and the absolute gain calibration to be applied to each array-band. Unfortunately, the dataset lacks the sensitivity to estimate the effective beam for polarization, which should not include a transfer function \citep{Naess_2023, Naess_DR6_2025} and might further need a polarization efficiency correction. Therefore, since these beams are approximate and may not be accurate at low multipoles,\footnote{The ACT DR6 daytime beams were published with the following caveats: ``The beams are only applicable to Stokes I because they include the low-multipole map-making transfer function that is present in the Stokes I maps but not in the polarization maps. The beams should work decently for polarization at multipoles above $\ell \sim 1000$, but users should be careful and check that results are stable with choice of $\ell_\textrm{min}$'' (\href{https://lambda.gsfc.nasa.gov/product/act/act_dr6.02/act_dr6.02_harmonic_beams_profiles_info.html}{LAMBDA ACT (DR6.02) Harmonic Beam Profiles}).} 
{we taper them to unity below $\ell=100$} and further run specific tests to ensure that our results are robust to the choice of the minimum multipole used in the lensing power spectrum reconstruction, which we describe in more detail in the next section.\medskip

After pre-processing the maps as described above, we ensure that the temperature power spectrum for each resulting array-band map is consistent between night and day. We do this by measuring the relative power spectrum of day with respect to night in the range $1000<\ell<2000$. We find that this ratio is constant within that range, and treat this factor as a gain calibration. The factors are reported in Table \ref{tab:cals}.\medskip

\begin{figure}
\includegraphics[width=0.85\columnwidth, trim=4mm 3mm 2mm 2mm, clip]
{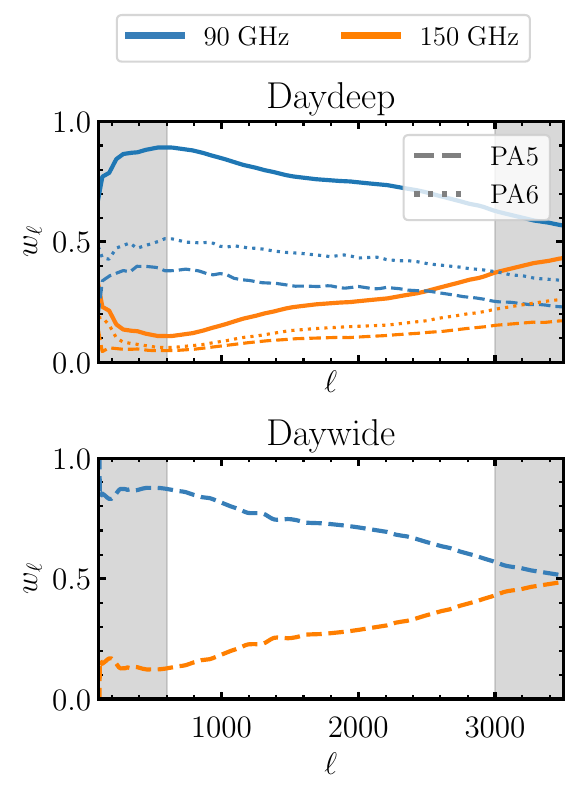}
\caption{\label{fig:weights_daytime} Maps from detector array-bands (PA) are combined using inverse-variance weights in harmonic space to form coadded \textit{daydeep} and \textit{daywide} sky maps, following the same approach as \citep{Frank_ACT_lensing_2024}. This figure shows the weights $w_\ell$ applied to each array-band map as a function of multipole. They sum to unity at each multipole to preserve the CMB signal. For \textit{daydeep}, the total weights assigned to each frequency are shown with solid lines. We shade in gray the CMB scales excluded in our analysis (for which we only include $600< \ell<3000$, informed by foreground contamination studies \citep{MacCrann_2024, AbrilCabezas_2025}). The PA6 \textit{daywide} dataset is excluded from our analysis as it is too shallow.}
\end{figure}

Finally, the different array-bands within each daytime dataset (\textit{daydeep}, \textit{daywide}) are combined in harmonic space using inverse-variance weighting following \citep{Frank_ACT_lensing_2024}, where the noise power spectra are estimated from the corresponding data splits \citep{Atkins_2023, Frank_ACT_lensing_2024}. The resulting weights are shown in Fig.~\ref{fig:weights_daytime}. The contribution from the \SI{150}{\giga\hertz} channels increases with multipole $\ell$ as the atmospheric noise (which scales with frequency) becomes subdominant. \medskip

\section{\label{sec:nulls}Measurement validation: null and consistency tests}We first perform a suite of null tests to ensure the robustness of our measurement, which we present in more detail in Appendix \ref{app:tablenulls}. Our analysis followed a blinding policy where no comparisons with theory predictions were allowed until the null tests were passed. Prior to unblinding, we also ensured that none of our null tests lay outside the range $0.001<\textrm{PTE}<0.999$.\footnote{The probability to exceed (PTE) is the probability of obtaining a higher $\chi^2$ than what we actually obtain, given a distribution with the same number of degrees of freedom.}  
The null tests are grouped in two categories: map-level and bandpower-level null tests. For each category, we run tests that include all array-bands coadded data, as well as tests that only include a subset of those (PA5, PA6, \SI{90}{\giga\hertz} and \SI{150}{\giga\hertz}-only tests). \medskip

\begin{figure}
\includegraphics[width=0.95\columnwidth,trim=3mm 4mm 2mm 2mm, clip]
{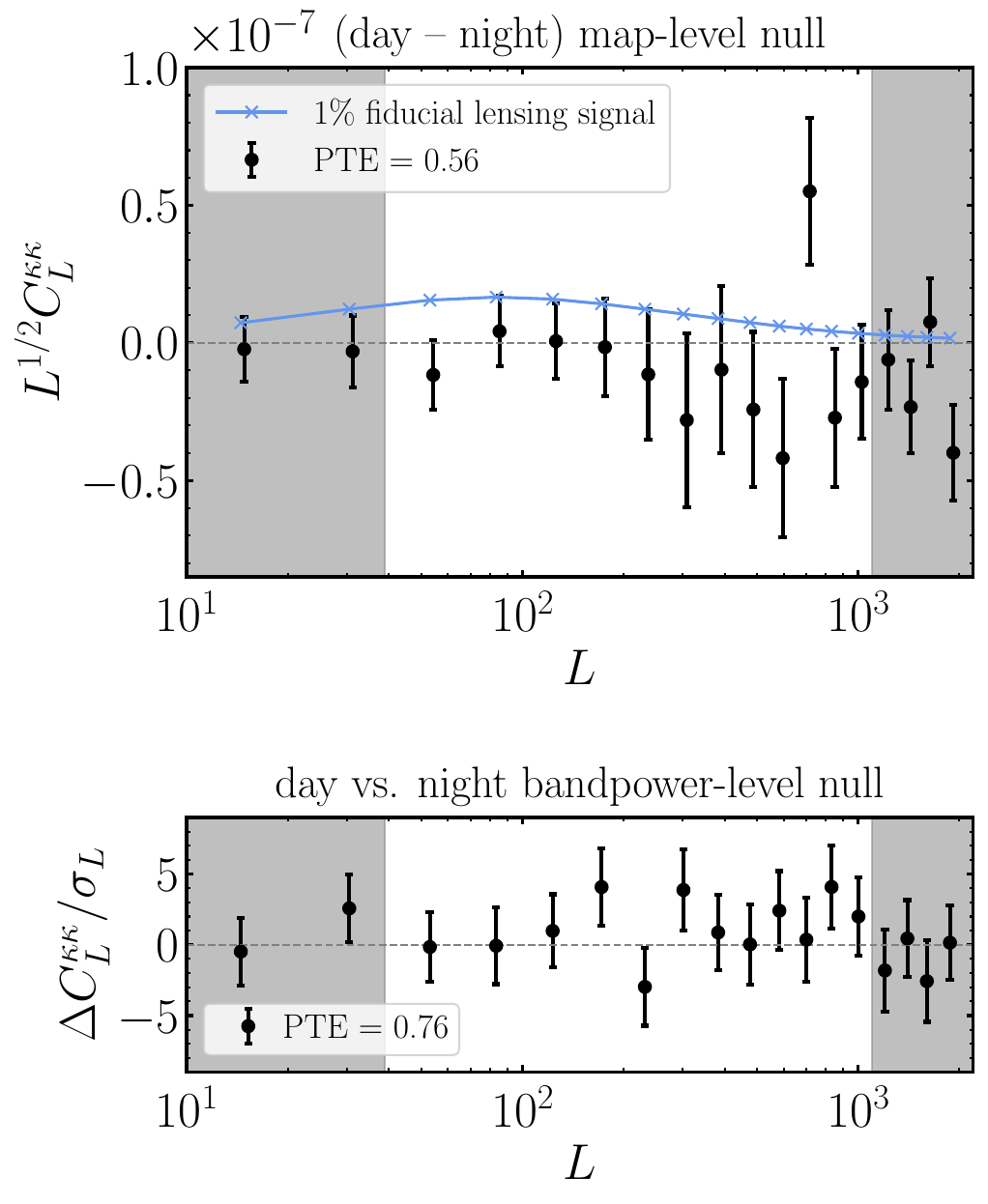}
\caption{\label{fig:maplevel} We validate the ACT DR6 \textit{daytime} lensing power spectrum with a suite of null tests targeted at potential systematics in the daytime data. \textit{Top}: The lensing power spectrum reconstructed from the difference between the daytime and nighttime maps. This is a stringent test: the signal is null within 1\% of the lensing signal (shown in blue). \textit{Bottom:} {Difference between the lensing power spectrum reconstructed from the daytime data and}
the ACT \texttt{dr6.02} nighttime-only lensing power spectra \citep{Kim_2026}, in units of the nighttime-only bandpower errors $\sigma_L$. The PTEs for these null tests are calculated from the goodness-of-fit with respect to null, by Monte Carlo sampling from the covariance matrix within the analysis range $40<L<1100$.
}
\end{figure}

\begin{figure}
\includegraphics[width=0.95\columnwidth,trim=3mm 3mm 2mm 2mm, clip]
{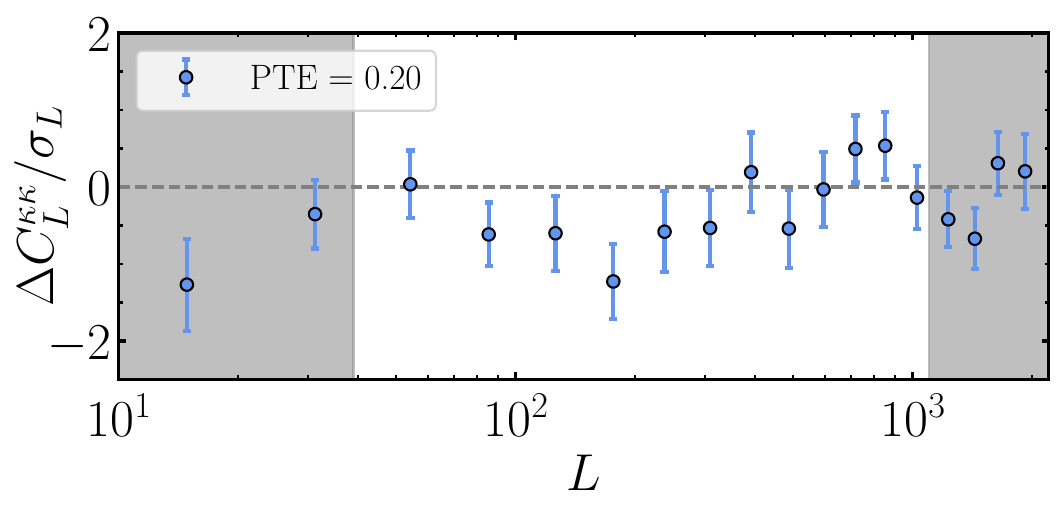}
\caption{\label{fig:lminpol1000} 
Difference in the reconstructed lensing bandpowers between our baseline analysis ($600<\ell_\textrm{CMB}<3000$) and a reconstruction where only $\ell_\textrm{CMB}>1000$ modes in polarization are used, divided by the error in the baseline bandpowers $\sigma_L$. This test is motivated by the fact that the daytime beams are approximate because the polarization beam includes a transfer function at $\ell_\textrm{CMB} < 1000$ that should only be present in temperature. We find consistency with null, with a PTE of 0.20. Excluding $\ell_\textrm{CMB}<1000$ modes in polarization minimally affects the sensitivity of our measurement (SNR of 17$\sigma$, c.f. baseline is 18$\sigma$). 
}
\end{figure}

\begin{figure}
\includegraphics[width=0.95\columnwidth,trim=2mm 3mm 2mm 2mm, clip]
{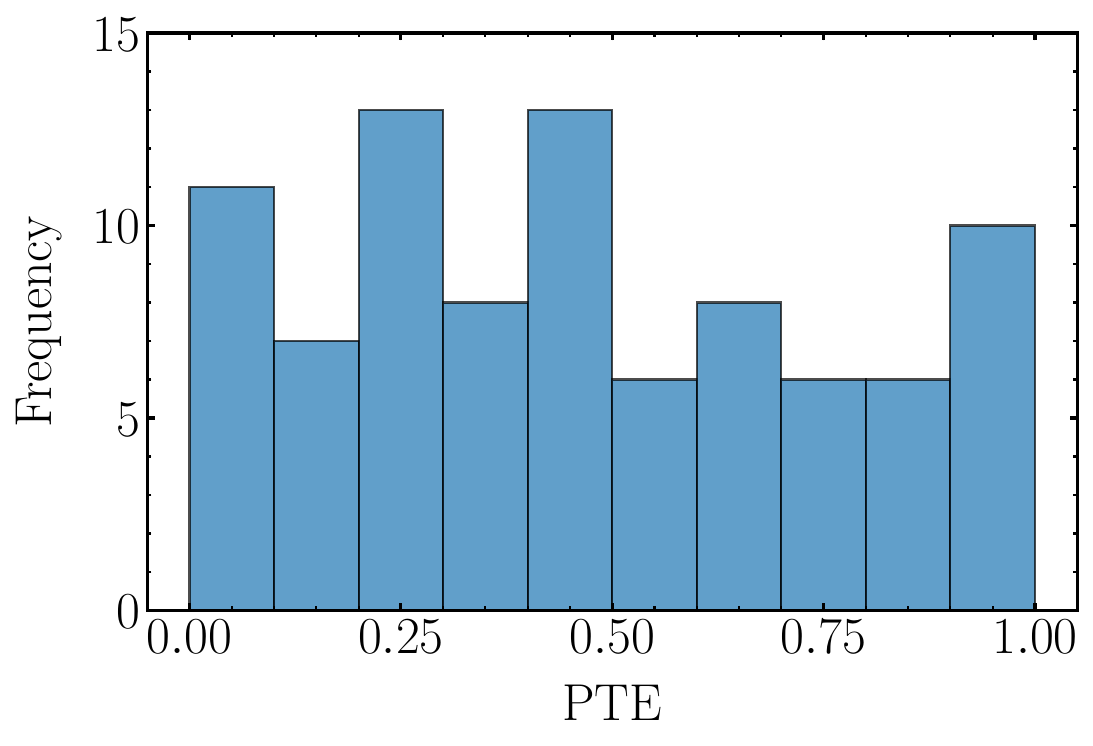}
\caption{\label{fig:hists} Histogram of PTE values obtained from all \textit{daydeep} and \textit{daywide} null tests. The distribution is consistent with a uniform distribution (Kolmogorov--Smirnov test PTE of 13\%; with the caveat that this test ignores correlations across different tests). None of our null tests lie outside the range $0.001<\textrm{PTE}<0.999$.}
\end{figure}

For map-level null tests, we perform the lensing reconstruction on maps that should contain no signal in the absence of systematic effects. For this, we prepare night vs.~day difference maps that are passed into the lensing reconstruction pipeline. We run the reconstructions using the minimum-variance temperature-polarization combination (MV), temperature-only (TT) and polarization-only (MVPol) quadratic estimators (see e.g., \citep{Maniyar_2021} for a review on lensing estimators). An example is shown in the top panel of Fig.~\ref{fig:maplevel}, where we show the lensing reconstruction using the MV estimator on the difference of day and night maps (all frequencies coadded). This is done for both \textit{daydeep}  and \textit{daywide} data, and the resulting null-test bandpowers are combined using inverse-variance weighting and a covariance derived from simulations. The difference is consistent with zero at high precision: since map-level nulls remove the CMB signal sample variance, the errors are remarkably small, making them extremely powerful null tests. In some cases, the errors are less than a hundredth of the signal power (for reference we show 1\% of the lensing signal in blue in Fig. \ref{fig:maplevel}). Therefore, these tests can thus strongly disfavor any additive systematic in the daytime maps that mimics lensing. We quantify the consistency with null by computing the $\chi^2$ with respect to null, which we translate into a PTE value by Monte Carlo sampling from the covariance matrix (PTE of 0.56 for the map-level test presented in Fig. \ref{fig:maplevel}). These are presented in Table \ref{tab:mapnull} (Appendix \ref{app:tablenulls}).\medskip

With bandpower-level null tests, we ensure that the lensing spectra using daytime data are consistent with those of night (Table \ref{tab:bandnull}, Appendix \ref{app:tablenulls}; see also bottom panel of Fig.~\ref{fig:maplevel}). In addition, we ensure internal consistency between our baseline MV reconstruction and those obtained when including less extended ranges of CMB modes (Table \ref{tab:bandnull}). Variations in the multipole range test for the impact of foregrounds and the presence of any ground pickup or any additional transfer function. In particular, the polarization beam is approximate, presenting a transfer function at low $\ell$ that should only be present in temperature.  Therefore, we ensure consistency between the bandpowers derived in our baseline analysis and a reconstruction where only $\ell_\textrm{CMB}>1000$ modes in polarization are used. The result is shown in Fig.~\ref{fig:lminpol1000}, showing that the difference between both is consistent with null (PTE of 0.41). Finally, we also check that the curl of the reconstructed deflection field is consistent with zero (Table \ref{tab:curl}).  Only systematic effects could induce a nonzero curl-like signal at these noise levels. Curl null tests are also an excellent test of our simulations and covariance matrices. \medskip

\begin{figure*}
\includegraphics[width=0.995\columnwidth]
{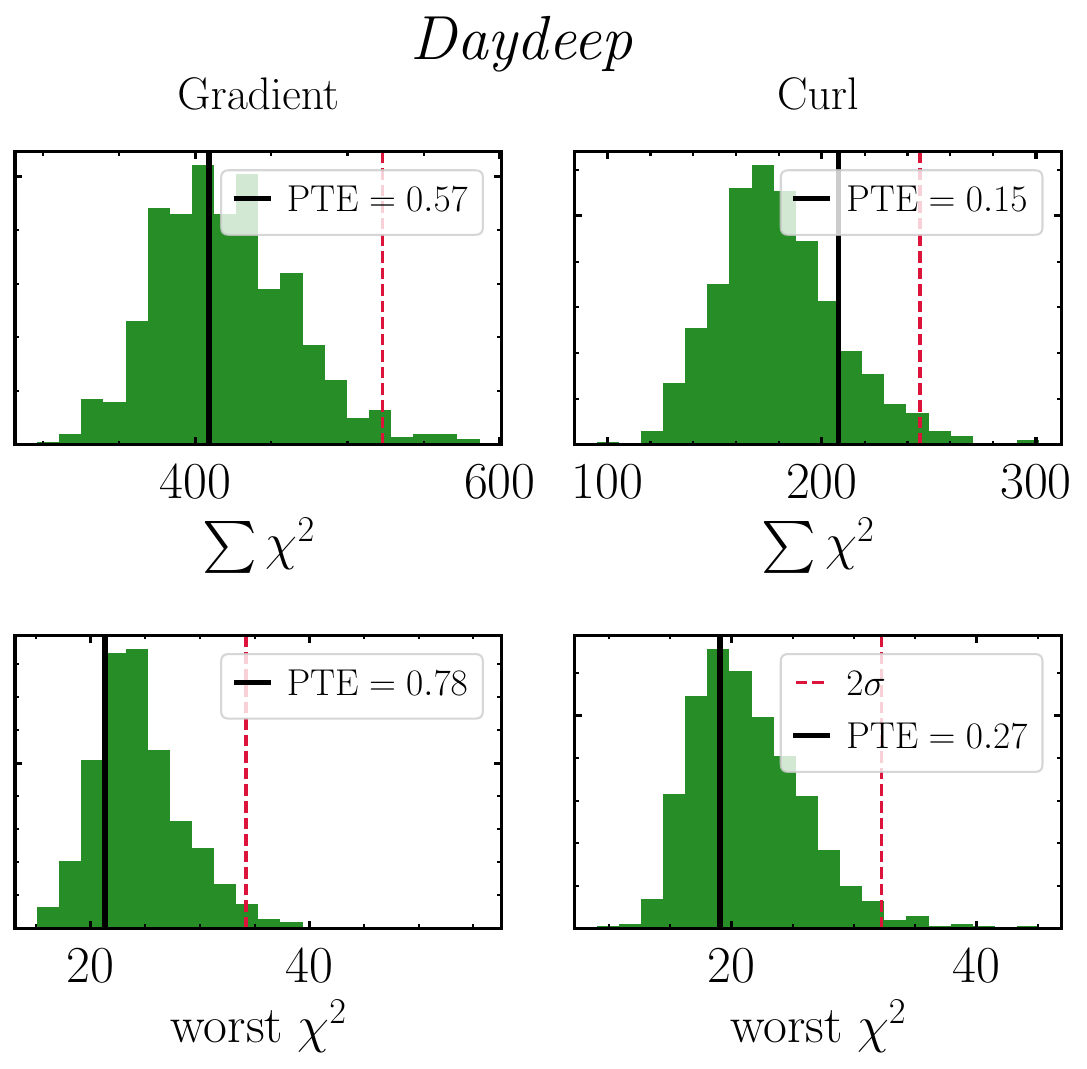}
\includegraphics[width=0.995\columnwidth]
{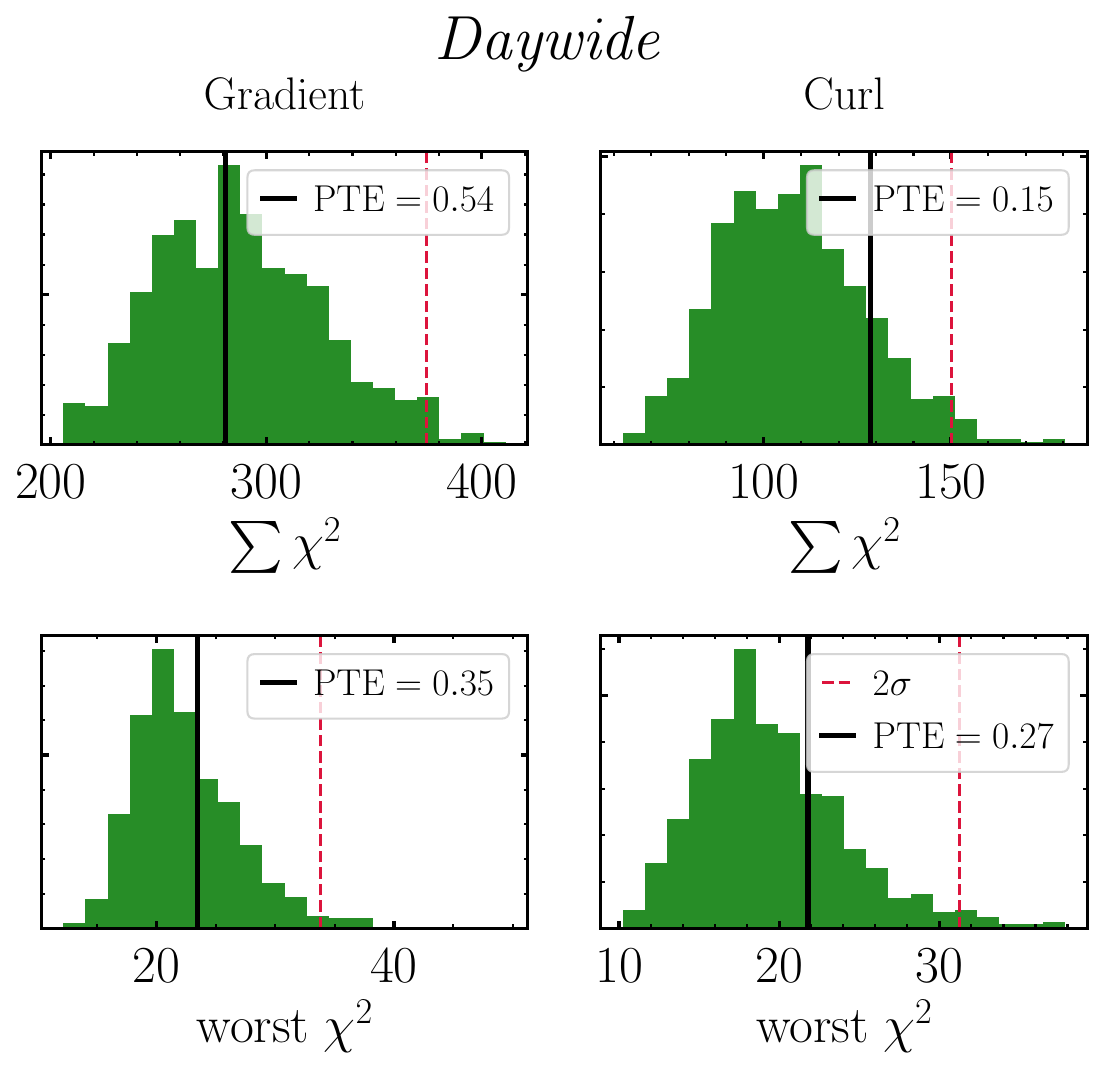}
\caption{\label{fig:pte_dist}{We examine our null-test suite as a whole by comparing the sum of $\chi^2$ (top row) and worst (bottom row) $\chi^2$ in the  data (solid black lines) to the distribution for the same summary statistic derived from simulations (green histograms). We do this for both the \textit{daydeep} (first two columns) and \textit{daywide} (last two columns) datasets, both for the gradient (columns 1, 3) and curl (columns 2, 4) of the deflection field. The PTE is then obtained as the fraction of simulations that exceed the data-derived values. For reference, we also show the simulation-derived $2\sigma$ (97.72 percentile) as a dashed red line. These tests do not reveal anything suspicious in our null-test suite.
\medskip}}
\end{figure*}

We summarize the PTE results of all \textit{daydeep} and \textit{daywide} null tests in Fig.~\ref{fig:hists}. Overall, the PTE distribution is consistent with a uniform one (Kolmogorov--Smirnov test PTE of 13\%; with the caveat that this test ignores correlations between the different tests). {We further study the overall null-test distribution for the \textit{daydeep} and \textit{daywide} datasets by comparing the sum of $\chi^2$ values and the worst (i.e. largest) $\chi^2$ found in the null-test suite to the distribution for the same summary statistic derived from simulations. We consider both the gradient and curl of the lensing deflection. The results are shown in Fig.~\ref{fig:pte_dist}, where the solid black lines represent the data-derived values and the green distributions are histograms of the values computed for every simulation. The PTE is then obtained as the fraction of simulations that exceed the data-derived values. For reference, we also show the simulation-derived $2\sigma$ (97.72 percentile) as a dashed red line. These tests naturally account for the correlation between the different tests, and do not reveal anything suspicious in our null-test suite.}\medskip

\medskip

\section{\label{sec:alens}Measurement of the lensing power spectrum}
\begin{figure*}
\includegraphics[width=0.95\textwidth, trim=2mm 2mm 2mm 2mm, clip]{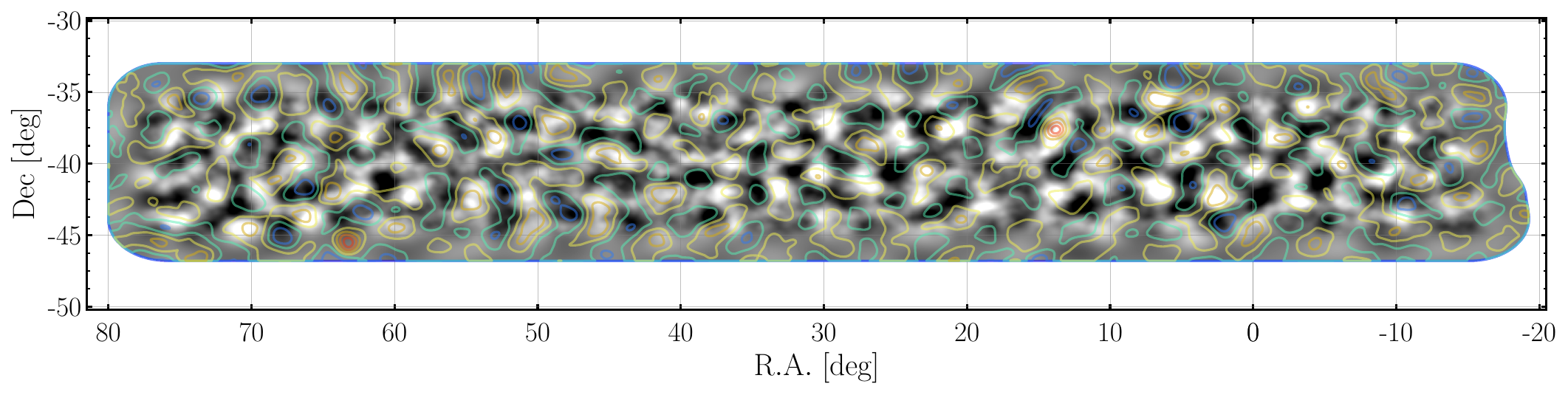}
\caption{\label{fig:daymap} Mapping the dark matter distribution during the day. Wiener-filtered map of the reconstructed projected matter density field in the southern \textit{daydeep} region (grayscale). Overlaid is a map of the CIB as measured by \textit{Planck} at \SI{545}{\giga\hertz} \citep{Planck_2016} (underdensities in blue, overdensities in red). Both signals should be highly correlated, based on their similar $z\sim2$ origin \citep{Song_2003}. Indeed, one can see by eye the correspondence between the two fields.}
\end{figure*}

We run the lensing reconstruction for both \textit{daydeep}  and \textit{daywide}, using the MV estimator with profile hardening \citep{Namikawa_2013, Osborne_2014, Sailer_2020, Sailer_2023} {to mitigate contamination from extragalactic foregrounds}. As an example, we show in Fig.~\ref{fig:daymap} shows the reconstructed projected matter density field in the southern \textit{daydeep} region {(\SI{1155}{\deg\squared})}. White regions correspond to peaks in the matter density field, while dark regions correspond to voids. One can see by eye the correspondence between the mass map and the Cosmic Infrared Background (CIB) measured by \textit{Planck} \citep{Planck_2016}, overlaid in contours (where blue and red correspond to underdensities and overdensities, respectively). These two fields are expected to be highly correlated \citep{Song_2003}.\medskip

\begin{figure*}
\includegraphics[width=0.75\textwidth,trim=3mm 3mm 2mm 1mm, clip]{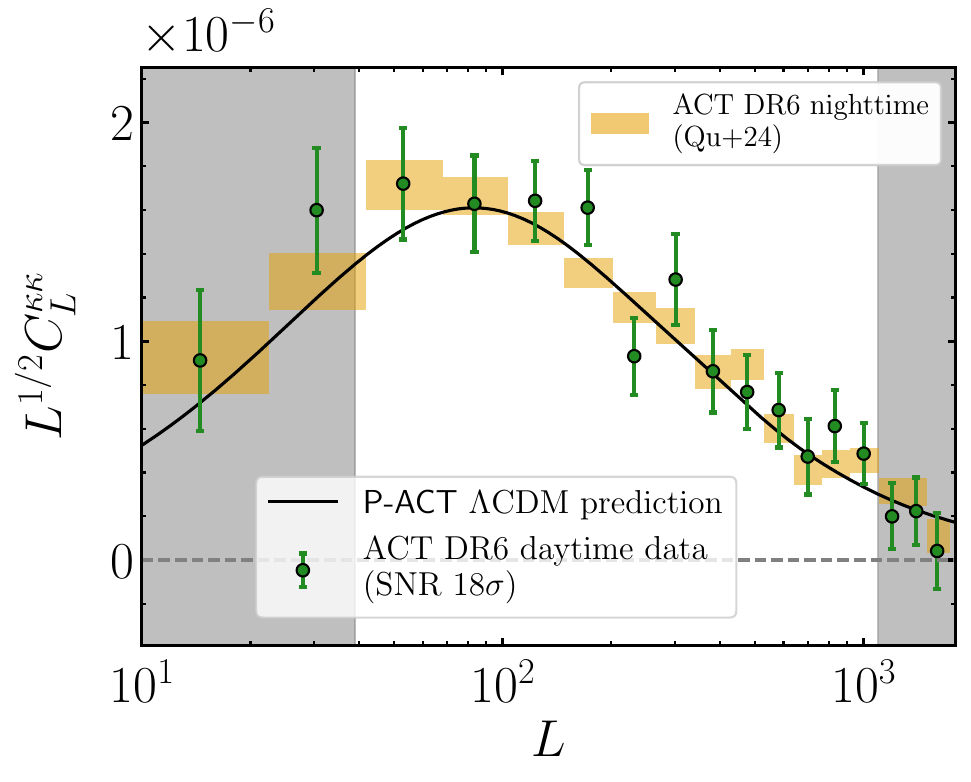}
\caption{\label{fig:daycmb} ACT DR6 \textit{daytime} lensing potential power spectrum, detected at $18\sigma$ within the analysis range $40<L<1100$ (unshaded region; choice verified in \citep{Frank_ACT_lensing_2024}, see main text for further details). The daytime spectrum results from combining the \textit{daydeep}  and \textit{daywide} lensing spectra (obtained from both temperature and polarization maps) at the bandpower level. We show in black the $\Lambda\textrm{CDM}$ prediction from the \textit{Planck}-ACT (\textsf{P-ACT}) CMB power spectrum  \citep{Naess_DR6_2025, Louis_DR6_2025, Calabrese_DR6_2025}. We find a lensing power spectrum amplitude {$A_\textrm{lens} = 1.071\pm0.060$ with respect to this model (PTE of 74\%)}. For reference, we show in boxes the current state-of-the-art ACT lensing analysis with nighttime data  (43$\sigma$ significance \citep{Frank_ACT_lensing_2024}).}
\end{figure*}

We combine the \textit{daydeep}  and \textit{daywide} lensing spectra at the power-spectrum level with a covariance derived from simulations. The resulting daytime-only lensing power spectrum $\hat{C}_L^{\kappa\kappa}$ is presented in Fig.~\ref{fig:daycmb} and Table~\ref{tab:clkk}, with a detection significance of $18\sigma$ within the {analysis range $40<L<1100$. We slightly expand the baseline multipole range chose in \citep{Planck_2020_lensing, Frank_ACT_lensing_2024} based on null-test stability and foreground estimates \citep{Frank_ACT_lensing_2024, MacCrann_2024, Kim_2026, AbrilCabezas_plus_2026}. Our binning minimizes the correlation between bandpowers: 12 bins with nonoverlapping bin edges at $[40,66,101,145,199,264,339,426,526,638,763, 902, 1100]$.}\medskip

The ACT DR6 \textit{daytime} lensing bandpowers are well fit by a lensing amplitude $A_\textrm{lens} = 1.071\pm0.060$ relative to the \textit{Planck}-ACT (\textsf{P-ACT}) \citep{Naess_DR6_2025, Louis_DR6_2025, Calabrese_DR6_2025} $\Lambda\textrm{CDM}$ prediction (PTE of 74\%), which is shown in black in Fig.~\ref{fig:daycmb}.\medskip

We use a Gaussian likelihood $\mathcal{L}$ to obtain constraints on cosmological parameters:\footnote{The likelihood is publicly available at \url{https://github.com/ACTCollaboration/dr6plus_lenslike}.} 
\begin{equation}
-2\ln{\mathcal{L}}\propto\sum_{bb^\prime}\big[\hat{C}^{\kappa\kappa}_{L_b}-{C}^{\kappa\kappa}_{L_b}(\boldsymbol{\theta})\big]{\mathbb{C}}^{-1}_{bb^\prime}\big[\hat{C}^{\kappa\kappa}_{L_{b^\prime}}-{C}^{\kappa\kappa}_{L_{b^\prime}}(\boldsymbol{\theta})\big],
 \end{equation}
where ${C}^{\kappa\kappa}_{L_b}(\boldsymbol{\theta})$ is the theory lensing spectrum for a given cosmological parameter combination $\boldsymbol{\theta}$, and ${\mathbb{C}}_{bb^\prime}$ is the covariance matrix {(computed from 764 simulations; we then rescale the inverse covariance by the Hartlap factor \citep{Hartlap_2007})}. We summarize our choice of priors in Table \ref{tab:priors}. These  are consistent with previous lensing analyses \citep{Planck_2020_lensing, Frank_ACT_lensing_2024, Madhavacheril_2024, Qu_2025}. \medskip

\begin{figure}
\includegraphics[width=0.998\columnwidth, trim=2mm 2mm 2mm 2mm, clip]{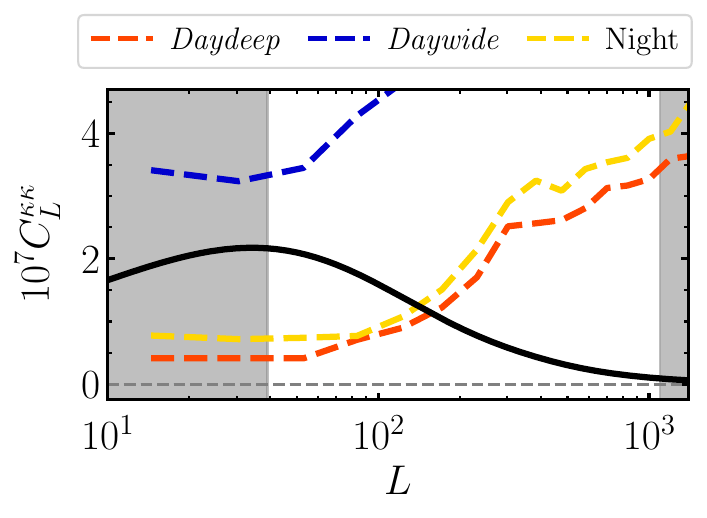}
\caption{\label{fig:noise} 
Variance of the reconstruction noise per-mode for the \textit{daydeep} (orange), \textit{daywide} (blue) and night (yellow) ACT \texttt{dr6.02} datasets, in comparison to the fiducial \textsf{P-ACT} lensing power spectrum prediction (shown in black).}
\vspace{-5mm}
\end{figure}

In combination with Baryon Acoustic Oscillation (BAO) measurements from the Dark Energy Spectroscopic Instrument (DESI DR2) \citep{DESI_DR2_2025a, DESI_DR2_2025b}, we constrain the amplitude of matter fluctuations $\sigma_8$
\begin{equation}
    \sigma_8 = 0.842 \pm 0.026 \quad (\textrm{ACT DR6 day + DESI DR2}).
\end{equation}
This 3\% measurement on $\sigma_8$ is consistent within less than $1\sigma$ with the value obtained by the ACT DR6 nighttime lensing analysis \citep{Madhavacheril_2024} and by the recent ACT, \textit{Planck} and SPT lensing combination \citep{Qu_2025}. \medskip

Fig.~\ref{fig:noise} presents the per-mode reconstruction noise power for the ACT \texttt{dr6.02} datasets.  Both night and \textit{daydeep} are signal-dominated on scales $L\lesssim 150$. While the \textit{daydeep} maps have a greater depth, they only cover a fifteenth of the night's area.  \textit{Daywide} has a reconstruction noise power more than three times larger than the other two datasets. {We defer a detailed discussion of the relative contributions of daytime and nighttime data to the final lensing power spectrum to the upcoming  ACT DR6+ lensing papers \citep{Kim_2026, AbrilCabezas_plus_2026, Quplus_2026}.}\medskip

\section{\label{sec:discussion}Conclusion}We have presented the first CMB lensing power spectrum analysis using data gathered by the Atacama Cosmology Telescope during the day. We detect the signal at {$18\sigma$} significance, and validate the resulting lensing power spectrum with a large suite of null tests that target different potential systematic effects in the daytime data and that also ensure consistency with the nighttime bandpowers. We anticipate including this dataset in the upcoming ACT DR6+ lensing analysis \citep{Kim_2026, AbrilCabezas_plus_2026, Quplus_2026}.\medskip

This is a proof-of-principle demonstration that lensing with daytime data is a possibility, and it is encouraging for future lensing measurements with the Simons Observatory (which is specifically designed to enable full daytime functionality)  \citep{SO_2019,ASO_2025}\footnote{The main difference with ACT is the use of a carbon fiber backup structure for the mirrors which keeps the mirror panels aligned under different thermal conditions. In addition, the mirrors are housed in a way that minimizes solar radiation further reducing daytime effects.} in order to place tight constraints on cosmological parameters.

\begin{acknowledgements}
We thank Bruce Partridge  for providing insightful comments on an early paper draft. I.A.C. acknowledges support from Fundaci\'on Mauricio y Carlota Botton and the Cambridge International Trust. B.D.S. acknowledges support from the European Research Council (ERC) under the European Union’s Horizon 2020 research and innovation program (Grant Agreement No. 851274). E.C. acknowledges support from the European Research Council (ERC) under the European Union’s Horizon 2020 research and innovation programme (Grant Agreement No. 849169). A.C. acknowledges support from the STFC (Grant No. ST/W000977/1 and ST/X006387/1). N.S. acknowledges support from DOE Award No. DE-SC0025309. C.S. acknowledges support from the Agencia Nacional de Investigaci\'on y Desarrollo (ANID) through Basal project FB210003. K.M. acknowledges support from the National Research Foundation of South Africa.  M.H. acknowledges support from the National Research Foundation of South Africa (Grants No. 97792, 137975, CPRR240513218388)
This work received support from the U.S. Department of Energy under Contract No. DE-AC02-76SF00515 to SLAC National Accelerator Laboratory.\medskip

This research has also made extensive use of \textsc{tempura},\footnote{\url{https://github.com/simonsobs/tempura}}
\textsc{falafel},\footnote{\url{https://github.com/simonsobs/falafel}}
\textsc{so-lenspipe},\footnote{{\url{https://github.com/simonsobs/so-lenspipe}}}
\textsc{orphics}\footnote{\url{https://github.com/msyriac/orphics}}
and the \textsc{mnms} \citep{Atkins_2023}, \textsc{astropy} \citep{astropy}, \textsc{numpy} \citep{numpy} and \textsc{scipy} \citep{scipy} packages. We also acknowledge use of the \textsc{matplotlib} \citep{matplotlib} package to produce the plots in this paper, use of the Boltzmann code \textsc{camb}~\citep{CAMB} for calculating theory spectra, and use of the \textsc{getdist} \citep{1910.13970}, \textsc{cobaya} \citep{2005.05290} and \textsc{cosmosis} \citep{1409.3409} software for likelihood analysis and sampling. 
This work was performed using resources provided by the Cambridge Service for Data Driven Discovery (CSD3) operated by the University of Cambridge Research Computing Service (\url{www.csd3.cam.ac.uk}), provided by Dell EMC and Intel using Tier-2 funding from the Engineering and Physical Sciences Research Council (Capital Grant EP/T022159/1) and DiRAC funding from the Science and Technology Facilities Council (\url{www.dirac.ac.uk}). In particular, this work used the DiRAC Data Intensive service. The DiRAC component of CSD3 at Cambridge was funded by BEIS, UKRI and STFC capital funding and STFC operations grants. DiRAC is part of the UKRI Digital Research Infrastructure.\medskip

We are grateful to the NASA LAMBDA archive for hosting our data. This work uses products from the Planck satellite, based on observations obtained with Planck (\url{http://www.esa.int/Planck}), an ESA science mission with instruments and contributions directly funded by ESA Member States, NASA, and Canada.\medskip

Support for ACT was through the U.S.~National Science Foundation through Awards No. AST-0408698, AST-0965625, and AST-1440226 for the ACT project, as well as Awards No. PHY-0355328, PHY-0855887 and PHY-1214379. Funding was also provided by Princeton University, the University of Pennsylvania, and a Canada Foundation for Innovation (CFI) award to UBC. ACT operated in the Parque Astron\'omico Atacama in northern Chile under the auspices of the Agencia Nacional de Investigaci\'on y Desarrollo (ANID). The development of multichroic detectors and lenses was supported by NASA Grants No. NNX13AE56G and NNX14AB58G. Detector research at NIST was supported by the NIST Innovations in Measurement Science program. Computing for ACT was performed using the Princeton Research Computing resources at Princeton University, the National Energy Research Scientific Computing Center (NERSC), and the Niagara supercomputer at the SciNet HPC Consortium. SciNet is funded by the CFI under the auspices of Compute Canada, the Government of Ontario, the Ontario Research Fund–Research Excellence, and the University of Toronto. We thank the Republic of Chile for hosting ACT in the northern Atacama, and the local indigenous Licanantay communities whom we follow in observing and learning from the night sky.\medskip

\end{acknowledgements}

\section*{Data availability}

The ACT DR6 dataset is publicly available on the Legacy Archive for Microwave Background Data Analysis (LAMBDA) at \url{https://lambda.gsfc.nasa.gov/ product/act/actadv_prod_table.html} and at NERSC at \texttt{/global/cfs/cdirs/cmb/data/act\_dr6/dr6.02}. The ACT DR6 daytime CMB lensing bandpowers, associated covariance and likelihood can be found at \url{https://github.com/ACTCollaboration/dr6plus_lenslike} and will be made available on LAMBDA.

\bibliographystyle{act_titles}
\bibliography{main}

\appendix

\begin{table}
\caption{\label{tab:mapnull}
Summary of PTE values inferred from the different map-level null tests, where the reconstruction was performed with the MV estimator. They show no evidence of an additive systematic in the daytime maps.
}
\begin{ruledtabular}
\begin{tabular}{cccc}
\textrm{Map-level null test}&
\textrm{\textit{Daydeep} }&
\textrm{\textit{Daywide}}&
\textrm{Daytime}\\
\colrule
coadd & 0.55 & 0.44 & 0.56 \\ [2pt]
\SI{90}{\giga\hertz} & 0.66 &  0.77 & 0.67 \\ [2pt]
\SI{150}{\giga\hertz} &  0.51 & 0.27 & 0.39 \\[2pt]
PA5 & 0.11 & 0.62 & 0.21 \\ [2pt]
PA6 & 0.64 & -- & -- \\[2pt]
\end{tabular}
\end{ruledtabular}
\end{table}

\begin{table}
\caption{\label{tab:bandnull}
Summary of PTE values derived from the different bandpower-level null tests. Overall, we find good agreement between the nighttime and daytime bandpowers. We also check that the daytime reconstructed bandpowers are stable with respect to using a more restricted range of CMB modes. }
\begin{ruledtabular}
\begin{tabular}{cccc}
\textrm{Bandpower-level null test}&
\textrm{\textit{Daydeep} }&
\textrm{\textit{Daywide}}&
\textrm{Daytime}\\
\colrule
coadd & 0.79 & 0.65 & 0.76 \\ [2pt]
\SI{90}{\giga\hertz} & 0.85 &  0.41 & 0.80 \\ [2pt]
\SI{150}{\giga\hertz} &  0.05 & \bf{0.024} & 0.04  \\[2pt]
PA5 & 0.44 & 0.53 & 0.10 \\ [2pt]
PA6 & 0.66 & -- & -- \\[2pt]
$\ell_\textrm{min}\textrm{(pol)}~1000$ & 0.41 & 0.79 & 0.20\\[2pt]
$1000<\ell_\textrm{CMB}<3000$ & 0.06 & 0.28 & 0.05\\[2pt]
$800<\ell_\textrm{CMB}<3000$  & 0.90 & 0.06 & 0.64\\[2pt]
$600<\ell_\textrm{CMB}<2000$  & 0.67 & 0.26 & 0.38\\[2pt]
$600<\ell_\textrm{CMB}<2500$  & 0.44 &  0.86 & 0.68\\[2pt]
\end{tabular}
\end{ruledtabular}
\end{table}

\begin{table}
\caption{\label{tab:curl}
PTE values for the different bandpower-level curl null tests. The curl of the reconstructed lensing deflection field is expected to be zero at these noise levels, and it is thus an excellent test of our simulations and covariance matrices.}
\begin{ruledtabular}
\begin{tabular}{cccc}
\textrm{Bandpower-level null test}&
\textrm{\textit{Daydeep}}&
\textrm{\textit{Daywide}}&
\textrm{Daytime}\\
\colrule
coadd & 0.39 & 0.51 & 0.32 \\ [2pt]
\SI{90}{\giga\hertz} & 0.88 &  0.22 & 0.77 \\ [2pt]
\SI{150}{\giga\hertz} & 0.38 & 0.04 & 0.34 \\[2pt]
PA5 & 0.11 & 0.51\footnotemark[1] & 0.05 \\ [2pt]
PA6 & 0.45 & -- & 0.45\footnotemark[2] \\[2pt]
\end{tabular}
\end{ruledtabular}
\footnotetext[1]{The \textit{daywide} result for PA5 (curl) is the same as all array-bands coadded (``coadd''), since this dataset does not include PA6. We do not double count this result in the histogram compilation.}
\footnotetext[2]{The ``Daytime'' result for PA6 (curl) is that of \textit{daydeep} PA6 (curl). This is only counted once in the histogram compilation.}
\end{table}

\begin{table}
\caption{\label{tab:clkk}ACT DR6 daytime CMB lensing bandpowers $\hat{C}_L^{\kappa\kappa}$.\protect\footnotemark[1] The multipole bin edges $[L_\textrm{min}, L_{\textrm{max}}]$ are given in the first column and the band centers $L_b$ in the second.}
\begin{ruledtabular}
    \begin{tabular}{ccc}
        $[L_{\text{min}},~{L}_{\text{max}}]$ & $L_b$ & $10^7\hat{C}^{{\kappa}{\kappa}}_{L}$ \\ \hline
$[40, 66]$                             & 53.0  & $2.36 \pm 0.35$                          \\
$[67, 101]$                           & 83.5  & $1.78 \pm 0.24$                          \\ 
$[102, 145]$                           & 123.0   & $1.48 \pm 0.17$                         \\ 
$[146, 199]$                           & 172.0   & $1.23 \pm 0.13$                          \\ 
$[200, 264]$                           & 231.5 & $0.61 \pm 0.12$                          \\ 
$[265, 339]$                           & 301.5 & $0.74 \pm 0.12$                          \\ 
$[340, 426]$                           & 382.5 & $0.441 \pm 0.096$                         \\ 
$[427 , 526]$                           & 476.0   & $0.352 \pm 0.078$                         \\
$[527 , 638]$                           & 582.0   & $0.284 \pm 0.071$                         \\
$[639, 763]$                           & 700.5 &$0.179 \pm 0.066$  \\
$[764, 902]$                           & 832.5 &$0.212 \pm 0.057$  \\
$[903, 1100]$                           & 1001.0 &$0.154 \pm 0.044$  \\
\end{tabular}
    \end{ruledtabular}
\footnotetext[1]{The datapoints and associated covariance $\mathbb{C}$ are available at \url{https://github.com/ACTCollaboration/dr6plus_lenslike}.}
\end{table}

\section{\label{app:tablenulls}Summary of null tests}We summarize in Tables \ref{tab:mapnull}--\ref{tab:curl} the PTEs of all null tests derived using the MV estimator. For each category, we run both all frequency-coadded arrays (row labeled ``coadd'') and subsets of those: single polarized arrays (PA5, PA6) and single frequencies (\SI{90}{\giga\hertz} and \SI{150}{\giga\hertz}). Except for the multipole-range stability tests, we always compare nighttime to daytime data. We obtain the reconstructions on both \textit{daydeep}  and \textit{daywide} data (second and third columns). We then sum the resulting bandpowers with inverse variance weighting and a covariance derived from simulations to obtain the final daytime bandpowers (last column, labeled ``Daytime'').  Table \ref{tab:mapnull} summarizes the results corresponding to the map-level null tests. They show no evidence of an additive systematic in the daytime maps. Table \ref{tab:bandnull} reports the PTEs for the corresponding bandpower-level null tests. {We find just one marginal failure,} when we compare the \textit{daywide} and nighttime \SI{150}{\giga\hertz} bandpowers.\medskip

We also test for consistency between our baseline bandpowers and those obtained when including less extended ranges of CMB modes, which we report in the {five} bottom rows of Table \ref{tab:bandnull}. When we check for stability using only $\ell_\textrm{CMB}>1000$ modes in polarization {(sixth row)}, we find good agreement with our baseline bandpowers. The latter test ensures that the {inclusion} of a transfer function within the polarization effective beam, which should really only be present in {temperature}, is not biasing our results in any {significant} way. Finally, we ensure that we detect no curl in the CMB lensing reconstructions above. These are reported in Table \ref{tab:curl}. Summary statistics for the overall null-test suite distribution regarding the curl of the lensing field are shown in the second and fourth column of Fig.~\ref{fig:pte_dist}, where we assure ourselves of the suitability of our simulations and covariance matrices (curl nulls are consistent with zero).\medskip

\section{\label{app:bandpowers}Daytime lensing bandpowers and cosmological inference}Complementary to Fig.~\ref{fig:daycmb}, Table \ref{tab:clkk} reports the ACT DR6 daytime CMB lensing bandpowers $\hat{C}^{\kappa\kappa}_{L}$. When we derive constraints on cosmological parameters, we impose priors which are summarized in Table \ref{tab:priors}. CMB lensing is not sensitive to the CMB optical depth (as we probe the total matter power spectrum), so it is fixed to $\tau=0.055$. The prior on $\Omega_bh^2$ is informed by the Big Bang nucleosynthesis prediction that matches the primordial deuterium abundance measurement \citep{Cooke_2018, Mossa_2020}.

\begin{table}
\caption{\label{tab:priors}Priors used to obtain the cosmological constraints presented in this work. Uniform priors are shown in square brackets and Gaussian priors with mean $\mu$ and standard deviation $\sigma$ are denoted $\mathcal{N}(\mu,\sigma)$.}
\begin{ruledtabular}
    \begin{tabular}{cc}
      Parameter       & Prior      \\ \hline
$\ln (10^{10}A_s)$ & $[2, 4]$           \\ 
$H_0 [\hun]$           & $[40,100] $        \\ 
$n_s$           & $\mathcal{N}(0.96,0.02)$     \\ 
$\Omega_bh^2$   & $\mathcal{N}(0.0223,0.0005)$ \\ 
$\Omega_ch^2$   & $[0.005,0.99]$    \\ 
$\tau$          & $0.055$  (fixed) 
                        \end{tabular}
    \end{ruledtabular}
\end{table}

\end{document}